\documentstyle[prl,aps,multicol,epsf]{revtex}

\begin{document}

\draft

\newcommand{\beq}{\begin{equation}}
\newcommand{\eeq}{  \end{equation}}
\newcommand{\bea}{\begin{eqnarray}}
\newcommand{\eea}{  \end{eqnarray}}
\newcommand{\bit}{\begin{itemize}}
\newcommand{\eit}{  \end{itemize}}

\title{Scars of the Wigner function.}

\author{Fabricio Toscano$^{1, 2}$
\thanks{\rm e-mail: toscano@cbpf.br}, Marcus A. M. de Aguiar$^{3}$ 
and Alfredo M. Ozorio de Almeida$^{1}$}

\address{$^{1}$Centro Brasileiro de Pesquisas F\'{\i}sicas, \\
Rua Xavier Sigaud 150, 22290-180, RJ, Rio de Janeiro, Brazil \\
\vspace{0.3cm}
$^{2}$ Instituto de F\'{\i}sica, Universidade do Estado do
Rio de Janeiro, \\  R. S\~ao Francisco Xavier 524, 20559-900,
RJ, Rio de Janeiro, Brazil. \\  
\vspace{0.3cm}
$^{3}$ Instituto de F\'{\i}sica,``Gleb Wataghin''.
Universidade Estadual de Campinas, \\ 83-970, Campinas, SP,
Brazil.}

\date{\today}

\maketitle

\begin{abstract}
We propose a picture of Wigner function scars as a sequence of 
concentric rings along a two-dimensional surface inside a periodic
orbit. This is verified for a two-dimensional plane
that contains a classical orbit of a Hamiltonian system with two 
degrees of freedom. 
The orbit is hyperbolic and the classical Hamiltonian is
``softly chaotic'' at the energies considered.
The stationary wave functions are the familiar mixture of scarred and
random waves, but the spectral average of the Wigner functions
in part of the plane is nearly that of a harmonic oscillator 
and individual states are also remarkably regular.
These results are interpreted in terms of the semiclassical picture of
chords and centres, which leads to a qualitative explanation of the 
interference effects that are manifest in the other region of the
plane.
The qualitative picture is robust with respect to a canonical
transformation that bends the orbit plane.

\end{abstract}


\begin{multicols}{2}

\narrowtext

%
%
%
\section*{}

\indent

Sixteen years have passed since Heller \cite{heller} detected scars
of periodic orbits in individual eigenfunctions of chaotic systems.
Explanations in terms of wave packets \cite{heller} or the 
semiclassical Green function \cite{bogo,libroalf} do predict
an enhancement of intensity near the projection of a Bohr-quantized
periodic orbit. 
However, such theories apply to a collective superposition
of states near this quantization condition.
In spite of some quite striking visual evidence, the issue of 
scarring for individual states, is confused by the fact that different
branches of the same periodic orbit may overlap so that their 
contributions interfere in the position space and this must be 
superposed on the expected random wave background for the chaotic 
state.
This has led to the need to make quantitative assessments of
scarring strength \cite{kahe,ka}.
We here show that the phase space picture can provide sharper
qualitative evidence of the influence of a periodic orbit on its 
Wigner functions. 

The old conjecture of Berry and Voros \cite{berrvoros} that the
Wigner functions for eigenstates of chaotic Hamiltonians 
are concentrated on the corresponding classical energy shells is
widely disseminated \cite{gutzbook}, as it is compatible with 
Schnirelman's theorem \cite{schzcdv}.
Nonetheless, no such restriction arises in the more recent 
semiclassical theory for the mixture of these states over a narrow
energy window, {\it i.e} the spectral Wigner function 
\cite{alfrep,notasaula}.
Indeed, the computations presented in this letter show that
individual Wigner functions, as well as their energy average
can oscillate with large amplitudes deep inside the energy shell.

\begin{figure}
\setlength{\unitlength}{1cm}
\begin{picture}(0,6)(0,0)
\put(0.1,0){\epsfxsize=8cm\epsfbox[83 290 512 552]{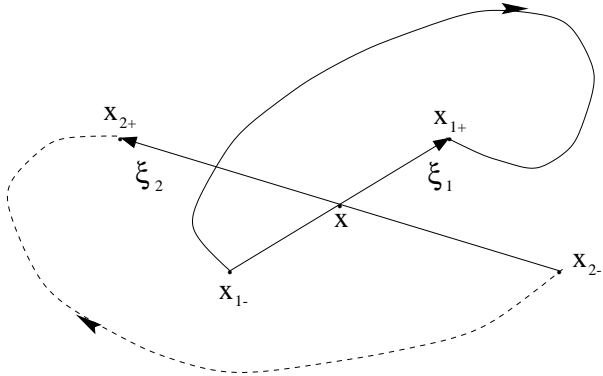}}
\end{picture} 
\vspace*{1.0pc}
\caption{\footnotesize In general there are many orbit segments
with endpoints ${\bf x}_{j-}$ and ${\bf x}_{j+}$ centered on a 
given point ${\bf x}$. The circuit, closed by the chord
$-\mbox{\boldmath $\xi$}_j={\bf x}_{j-}-{\bf x}_{j+}$, defines
the phase of the semiclassical contribution to 
(\ref{wigspecscl}).}
\label{Fig1}
\end{figure}

At points 
${\bf x}=({\bf q},{\bf p})$ inside
the energy shell, the semiclassical spectral Wigner function
is \cite{alfrep,notasaula}:
\beq
\label{wigspecscl}
W({\bf x},E,\varepsilon)=
\sum_j A_j({\bf x},E)\;e^{-\,\varepsilon t_j/\hbar}
\cos\left[\frac{S_j({\bf x},E)}{\hbar}+\gamma_j\right]
.
\eeq 
The sum is over all the trajectory segments on the $E$-shell with
endpoints, ${\bf x}_{j\pm}$, centered on ${\bf x}$, as sketched
in Fig.\ref{Fig1}.
The action is merely the symplectic area, 
$S_j=\oint {\bf p}\cdot d{\bf q}$, for the circuit taken
along the orbit from ${\bf x}_{j-}$ to ${\bf x}_{j+}$
and closed by the chord 
$-\mbox{\boldmath $\xi$}_j({\bf x})$.
The time of traversal for the stretch along the trajectory
is $t_j$ and $\varepsilon$ is the width of the energy window over 
which we average individual Wigner functions $W_n({\bf x})$,
{\it i.e.}, 
\beq
\label{wigspec}
W({\bf x},E,\varepsilon)=(2\pi\hbar)^{D/2}
\sum_n \frac{\varepsilon/\pi}{(E-E_n)^2+\varepsilon^2}\,
W_n({\bf x})
\;\;,
\eeq
where $D$ is the dimension of the phase space $({\bf q},{\bf p})$.
We shall not be concerned with the Maslov phase $\gamma_j$,
nor with the amplitude $A_j({\bf x},E)$,
except to note that these are purely classical quantities that 
vary smoothly with ${\bf x}$ inside the shell, as compared with 
the high frequency oscillations of the cosine factor.

%

The highly oscillatory nature of the spectral Wigner function inside
the energy shell is mostly washed out on projection, 
{\it i.e.} for true probability densities. 
However, the oscillations are unavoidable in the study of interference
effects. 

The energy shell itself is a caustic, because the chord 
$\mbox{\boldmath $\xi$} \rightarrow 0$ as ${\bf x}$ approaches the 
shell.
In this limit, the contributing trajectories either shrink to a point,
or they are very close to a periodic orbit. 
The modification of (\ref{wigspecscl}) leads to the Berry theory
of scars \cite{berry89}, futher refined in \cite{alfrep}, for 
evaluation points ${\bf x}$ close to the energy shell.

The point here is that large open segments of periodic orbits 
can be constructed by adding multiple windings to a primitive,
small segment of a periodic orbit.
If conditions for phase coherence, to be discussed, are satisfied,
we can then obtain a scar deep inside the energy shell.
This scar is located at the two-dimensional {\it central surface} 
constructed by the centers of all the chords with endpoints 
on the periodic orbit.
For the case where the phase space dimension $D=2$, the central surface 
is the phase space region enclosed by the convex hull of
the orbit that here is the energy shell 
\footnote{Note that if the orbit is convex the central surface is just 
the phase space region inside it.}.
For $D \geq 4$, the two-dimensional central surfaces are still
bounded by one-dimensional periodic orbits that lie within the 
higher dimensional energy shell.
In the simplest case where the orbit is plane and convex,
the central surface coincides with the part of the plane within
the orbit just as in the case where $D=2$.
If the orbit is not plane, then the central surface will be more
complicated and it may exhibit singularities and self-intersections.
In general it will not coincide everywhere with the invariant surface
formed by changing continuosly the energy of this periodic orbit.

We emphasize that these scars of the Wigner function that correspond to a very 
recognizable oscillatory pattern have special localization properties 
in phase space of dimension $D \geq 4$.
Previous explorations of phase space representations of 
eigenstates have been too blunt to discern these fine features.
The work of Feingold {\it et. al.} \cite{feingold} involves 
a projection from a three-dimensional section, which is fine for 
the Husimi function, but washes away the details of an oscillatory
Wigner function.
On the other hand, Agam and Fishman \cite{agamfishman} restrict their
investigation to the neighbourhood of the orbit and, hence, to the
energy shell.
This is just the edge of the central surface.

In this letter we tested our prediction for the simplest case of
an unstable periodic orbit lying in a two dimensional plane.
Accidentally this plane is a symmetry plane, however,
we have checked that the enhanced amplitude of the spectral 
Wigner function accompanies the distortion of the orbit due
to a nonlinear canonical transformation. 
Since the Wigner function is not invariant in this case,
we thus obtain an essentially new system, in which the new central
surface is not a symmetry surface.
Thus, our scarring effect cannot be attributed to a special
property of the symmetry plane.
This case is essentially different from recent studies of 
higher dimensional systems where the scars arise over marginally 
stable invariant planes that are indeed special \cite{prosen}.

\begin{figure}
\setlength{\unitlength}{1cm}
\begin{picture}(0,5)(0,0)
\put(0.1,0){\epsfxsize=8cm\epsfbox[48 264 548 578]{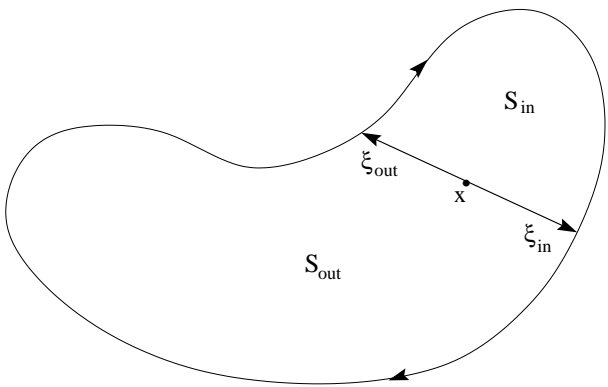}}
\end{picture}  
\vspace*{1.0pc}
\caption{\footnotesize If the endpoints lie on a periodic orbit,
we can add an infinite number of windings to the two shortest
primitive segments, ``{\it in}'' and ``{\it out}'', divided by 
the chords 
$\mbox{\boldmath $\xi$}_{in}=-\mbox{\boldmath $\xi$}_{out}$.
All the points ${\bf x}$, centers of chords in the periodic 
orbit, form a two dimensional {\it central surface}.
If the orbit is plane (as in the graph) the central 
surface is the part of the plane enclosed by the convex hull
of the orbit. 
}
\label{Fig2}
\end{figure}

The chord structure for segments of a periodic orbit is sketched
in Fig.\ref{Fig2}.
This is similar to a system with $D=2$,
for which Berry \cite{berry77} showed that there is only one 
chord for most points ${\bf x}$.
However, we must now distinguish between the chords 
$\mbox{\boldmath $\xi$}_{in}=-\mbox{\boldmath $\xi$}_{out}$ and 
the sum in (\ref{wigspecscl}) includes each different 
winding of the periodic orbit for primitive orbit segments,
``{\it in}'' and ``{\it out}'', associated with the two respective
chords. 
Evidently all these ``{\it in}'' and ``{\it out}'' contributions
build up if the energy is close to one of the values for 
which this periodic orbit is Bohr-quantized. 
We show in \cite{future} that the condition for these two 
contributions to be in phase is the same as that for a maximal 
contribution to the Gutzwiller trace formula \cite{gutzbook}.
Therefore, near the quantization energy, the sum of all the 
contributions in (\ref{wigspecscl}), owing to chords in the periodic
orbit, can be approximated by an expression analogous to Berry's
simplest semiclassical aproximation \cite{berry77,toscano} for the  
Wigner function in $D=2$ systems, but now evaluated over points 
${\bf x}$ in the two-dimensional central surface.
Hence, in this case, we have an enhanced contribution to the 
spectral Wigner function with the phase $S_{in}({\bf x})/\hbar$
and the Maslov phase $\gamma=0$ . 
The successive contours $S_{in}({\bf x})=$constant thus determine
rings of constant phase along the central surface.
Therefore, we predict rings of positive and negative amplitude
superimposed on the background of contributions from uncorrelated 
trajectory segments that also contribute to the spectral Wigner 
function.
The edge of this system of rings along the orbit itself is the object 
of the Berry theory \cite{berry89}, though it does not deal with
multiple windings.

We have tested this prediction for the coupled nonlinear 
oscillator ($D=4$)
\beq
\label{hamilnelson}
\mbox{H}({\bf q},{\bf p})=
\frac{(p_1^2+p_2^2)}{2} + 0.05\,q_1^2+
\left(q_2-\frac{q_1^2}{2}\right)^2
\;,
\eeq
sometimes referred to as the ``Nelson Hamiltonian''.
The classical periodic orbits together with the topology
of their families in the energy.vs.period plot has been studied in
\cite{baradavi}  and the wave functions and their scars were 
discussed in \cite{baraprovo}.
Evidently the plane $q_1=p_1=0$ is classically invariant, so
we obtain an isolated periodic orbit on this plane for each energy
(the vertical family).
This is then, the simplest case where the central surface is flat -
the central plane -.
The restriction of the Hamiltonian to this plane defines a
harmonic oscillator, so there is a single chord with endpoints
in the periodic orbit (except for sign) for
each point $(p_2,q_2)$ inside the orbit, except at $q_2=p_2=0$.

We computed the wave functions $\langle {\bf q}|n \rangle$ for the 
eigenstates of (\ref{hamilnelson}) within the range of energies
$0.821\leq E \leq 0.836$ taking $\hbar=0.05$ and using a basis of
eigenfunctions 
$\phi^{(q_1)}_{n_1}(q_1)\phi^{(q_2)}_{n_2}(q_2-q_1^2/2)$ ,
in the same way as in \cite{baraprovo}.
The Wigner function for each state was then calculated directly
by the double integral over ${\bf Q}=(Q_1,Q_2)$:
\beq
\label{wignerqpn}
W_n({\bf q},{\bf p})=\frac{1}{(2\pi\hbar)^2}
\int d{\bf Q}\, 
\langle {\bf q}+{\bf Q}/{2} |n\rangle 
\langle n|{\bf q}-{\bf Q}/{2} \rangle 
\,
e^{-i\frac{{\bf p}\cdot{\bf Q}}{\hbar}}
\, .
\eeq

\begin{figure} 
\setlength{\unitlength}{1cm}
\begin{picture}(0,11)(0,0)
\put(0.1,0){\epsfxsize=7.5cm\epsfbox[88 186 469 758]{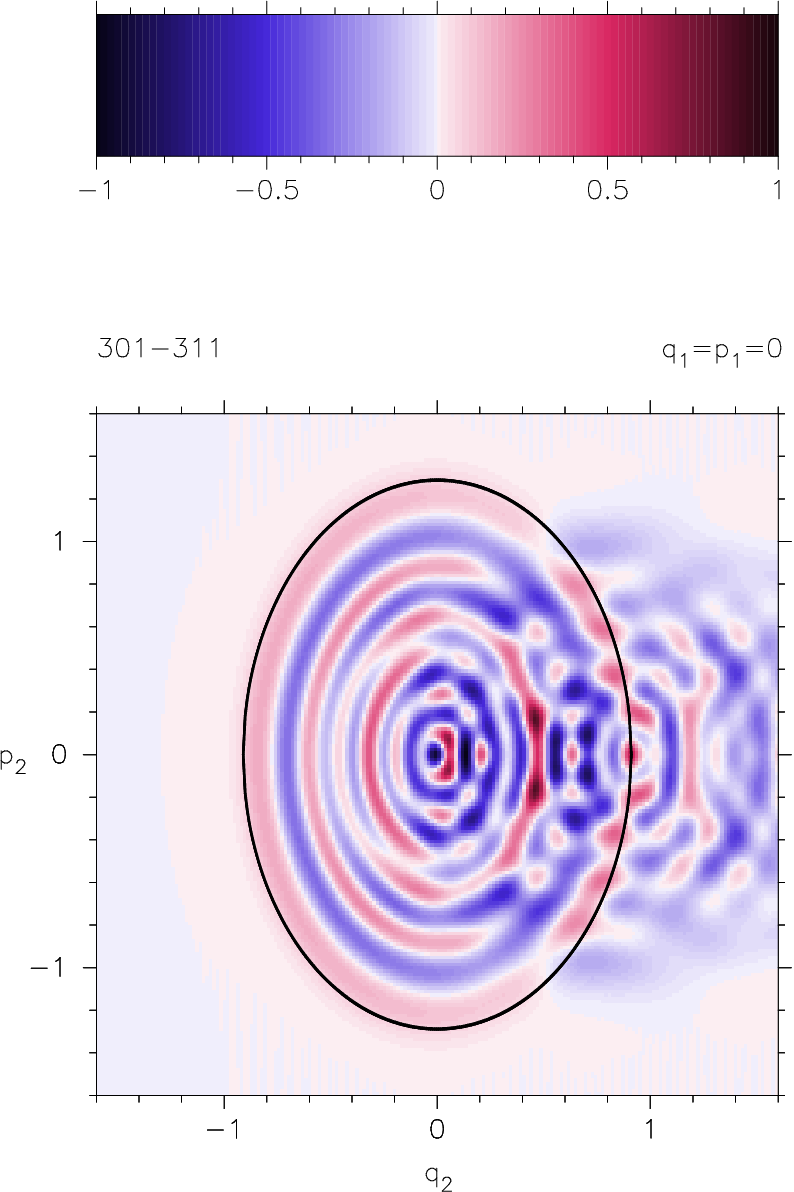}}
\end{picture} 
\vspace*{1.0pc}
\caption{\footnotesize
Averaged Wigner function on the $(q_2,p_2)$ plane for an energy
window containg eleven eigenstates centered on the $11$'th
Bohr level.
The thick dark line is the superposition of each energy shell in the
plane (periodic orbit) for energies in the range considered; the
middle one corresponding to the Bohr-quantization energy.}
\label{Fig3}
\end{figure}

In Fig.\ref{Fig3} we display the average over Wigner functions
for the energy window chosen, surrounding the 11'th Bohr energy 
for the vertical orbit that has Maslov index $3$.
This is more convenient than smoothing with the Lorentzian 
window in (\ref{wigspec}) and should produce essentially
the same results.

As an example,
Fig.\ref{Fig4} shows some of the individual Wigner functions in this
window for ${\bf x}$ in the $(q_2,p_2)$ plane. 
Immediatly, we notice that all these Wigner functions are remarkably
regular, roughly in the region $q_2<0$.
Indeed, Fig.\ref{Fig3} coincides in this region with the Wigner
function of the restricted Hamiltonian, also with respect to the 
precise phase of the oscillations. The individual Wigner functions 
follow the same phase contours, but there is a varying phase shift.
If we extrapolate the semiclassical theory to energy smoothings
of the order of the energy spacing, the conclusion is that there
are no orbit segments, other than the periodic orbit itself, 
contributing to the Wigner functions on this surface up to the 
Heisenberg time. 
This is more dramatic for the state $n=305$ whose
Wigner function on all the central surface is almost the same 
of that of the restricted Hamiltonian (see Fig.\ref{Fig4}).
It is important to note that only this state seems to have a 
visual scar of the vertical orbit in position space, although 
superimposed to a random wave background.  

For $q_2>0$, we have found other orbit segments that do not lie
on the central plane, but which do contribute to the spectral Wigner
function on it.
Indeed, this is the case for segments of both the symmetric periodic
orbits shown in Fig.\ref{Fig6} projected onto position space.
Needless to say, the multiple windings of these periodic orbits could
produce total contributions of the same order as the plane orbit, if
they are nearly Bohr-quantized. 
This supplies a qualitative explanation for the existence of  
interference patterns to the left of the Wigner functions 
in Fig.\ref{Fig3} and \ref{Fig4}.
Notice that the second of the orbits in Fig.\ref{Fig6} reaches into 
a region with $q_2$ greater than is attained by the plane periodic 
orbit.
So we account for Wigner functions reaching outside the energy
shell (which coincides with the periodic orbit on the central 
plane).
Note that these are only particular examples of the many symme\-tric
periodic orbits whose central surface intersects the central
plane along lines with $q_2>0$ \cite{baradavi,baraprovo}. 
Therefore, it will be hard to make quantitative predictions in this
region in contrast to that where $q_2<0$, for which no such orbits of
short period were found.
The enhanced scar pattern of the Wigner function of the state 
$n=305$ over all the central plane evidently washes out 
other possible contributions.   

The central surface of the harmonic orbit considered in this
work coincides with an invariant plane of the Hamiltonian 
(\ref{hamilnelson}) which is a reflection symmetry plane
of the system.
This might induce to the conclusion that the symmetry is the 
actual responsible for the scarring observed.
We show that this is not the case by applying a nonlinear 
canonical transformation, described in
\cite{future}, deforming the central plane $q_1=p_1=0$ into 
a new invariant surface (Fig.\ref{Fig5} {\bf a}).
Now, the new central surface of the distorted periodic orbit
no longer coincides everywhere with the invariant surface
(Fig.\ref{Fig5} {\bf b}) losing in this way any symmetry.
We calculated the Wigner functions for this new system,
over the central surface of the periodic orbit, for  
eigenstates within the same range of energies surrounding the
11'th  Bohr level. 
Both the average and the individual Wigner functions,  
projected over the $(q_2,p_2)$ plane, display the same features as 
those shown in Fig.\ref{Fig3} and Fig.\ref{Fig4}
respectively. 
Reference \cite{future} also considers the thickness of scar
surfaces.  

Our main point is that individual trajectory segments do play 
a role in the spectral Wigner function.
Indeed, phase coherent contributions from a periodic orbit can 
add up to scars in the form of concentric rings deep inside the 
energy shell.
Eventhough the Berry-Voros hypothesis \cite{berrvoros} ties
in more intuitively with the concept of quantum ergodicity,
it should be noted that the chord picture of Wigner functions
does not contradict Shnirelman's theorem and subsequent 
exact results \cite{schzcdv}, because the oscillations make 
negligible contributions to integrals with smooth functions.

It would seem that the structure of interfering chords,
from which the spectral Wigner function is built up, could only
generate a messy picture where scars would be even harder to 
recognize than in wave functions for which quantitative methods
are needed.
We have now shown that the converse can be true in the simplest case,
{\it i.e.} that the relatively low dimension of the central 
surface can allow it to miss, over an appreciable region, the
contribution of chords from other orbits. 
In this region we obtain a regular ring structure, not only
for the energy average, but even for the pure Wigner functions 
shown in Fig.\ref{Fig4}.
Though we cannot yet predict why the relative weight of this 
particular periodic orbit should vary so markedly among the states
close to the Bohr-quantized energy, we are now able to recognize
in Fig.\ref{Fig4} a remarkable example of an individual scar.

%
%

%
%

%
%

%
%
%
\begin{figure}
\setlength{\unitlength}{1cm}
\begin{picture}(0,8)(0,0)
\put(-0.2,5.0){\epsfxsize=8.5cm\epsfbox[25 81 539 332]{Fig4a.ps}}
\put(-0.2,0.0){\epsfxsize=8.5cm\epsfbox[25 81 539 332]{Fig4b.ps}}
\end{picture} 
\vspace*{1.0pc}
\caption{\footnotesize Some of the individual Wigner functions on the
$(q_2,p_2)$ plane that were averaged in Fig.\ref{Fig3}.
The plots are globally normalized to point out the enhanced
amplitude of the scar for the $n=305$ state.  
The eigenenergy of this state is at distance $\approx 1.5 \Delta$ 
from the Bohr energy ($\Delta$ the mean level spacing in the 
chosen average window).
The black ellipse, in each graph, is the periodic orbit 
for the corresponding eigenenergy.}
\label{Fig4}
\end{figure}

%
%
\begin{figure}
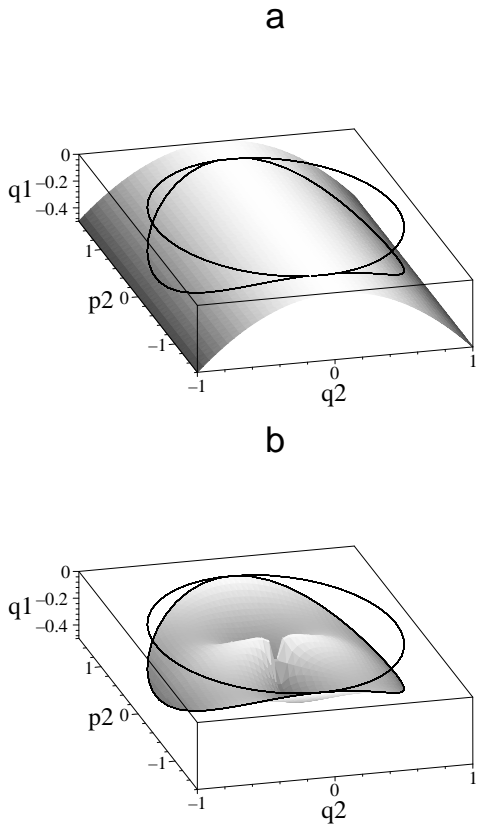

\setlength{\unitlength}{1cm}
\begin{picture}(8,9)(0,0)
\put(-1,5){\epsfxsize=9cm\epsfbox[80 265 469 527]{Fig5a.ps}}
\put(-1,-1.5){\epsfxsize=9cm\epsfbox[80 224 469 527]{Fig5b.ps}}
\end{picture} 
\vspace*{1.0pc}
\caption{\footnotesize The application of a particular nonlinear 
canonical transformation to the Hamiltonian (\ref{hamilnelson}) 
deforms the invariant plane $q_1=p_1=0$  into the surface 
displayed in the graph {\bf a}, that also has $p_1=0$.
The curve over this surface is the distortion of the original 
periodic orbit (ellipse displayed in the invariant plane).
The graph {\bf b} shows the new central surface for the deformed
periodic orbit.
}
\label{Fig5}
\end{figure}

%
%
\begin{figure}
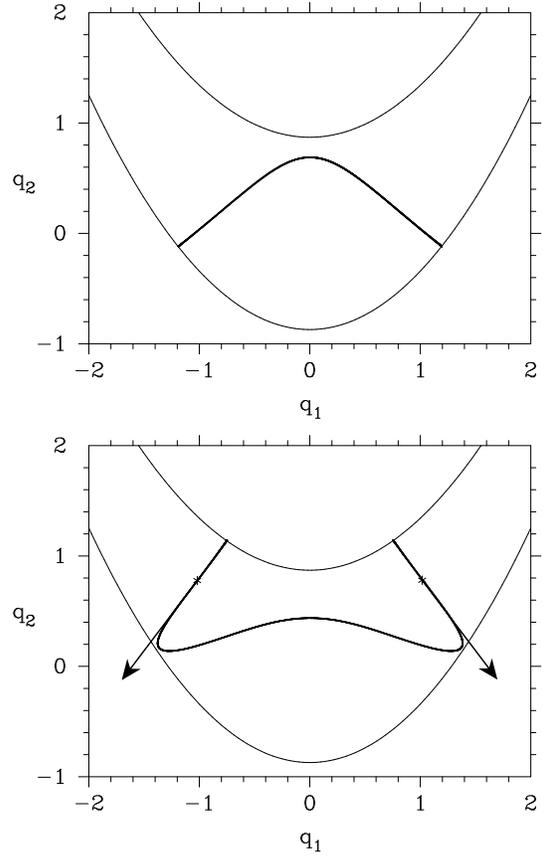

\setlength{\unitlength}{1cm}
\begin{picture}(0,11)(0,0)
\put(0.0,6.5){\epsfxsize=7cm\epsfbox[80 265 469 527]{Fig6a.ps}}
\put(0.0,0.0){\epsfxsize=7cm\epsfbox[80 224 469 527]{Fig6b.ps}}
\end{picture} 
\vspace*{1.0pc}
\caption{\footnotesize Two of the many symmetric periodic orbits 
that have chords centered on the $(q_2,p_2)$ plane. 
The graphs are in position space, so that the momenta for the 
symmetric endpoints are displayed as vectors. All centers have 
$q_2\stackrel{\scriptscriptstyle >}{\scriptscriptstyle \sim}0$.
The thin line is an equipotential of (\ref{hamilnelson}) for 
the energy of the $11$'th Bohr level.}
\label{Fig6}
\end{figure}

%
%

\acknowledgements

This work was financied by CNPq-CLAF, FAPERJ and Pronex-MCT.



\end{multicols}

\end{document}